# The collision frequency of electron-neutral-particle in the weakly ionized plasmas with Non-Maxwellian velocity distributions

Hong Wang, Jiulin Du, Rui Huo

*Department of Physics, School of Science, Tianjin University, Tianjin 300350, China*

**Abstract**

The collision frequencies of electron-neutral-particle in the weakly ionized complex plasmas with the non-Maxwellian velocity distributions are studied. The average collision frequencies of electron-neutral-particle in the plasmas are derived accurately. We find that these collision frequencies are significantly dependent on the power-law spectral indices of non-Maxwellian distribution functions and so they are generally different from the collision frequencies in the plasmas with a Maxwellian velocity distribution, which will affect the transport properties of the charged particles in the plasmas. Numerically analyses are made to show the roles of the spectral indices in the average collision frequencies respectively.

## 1. Introduction

Since many physical properties of plasmas depend on collision and collision frequency [1,2], the research on the influence of collision and collision frequency on plasma properties has attracted a large number of scientific researchers, such as Faussurier et al studied the plasma collision frequencies and the related plasma physics [3-5], Tsytovich *et al.* have discussed the effect of collisions between dust particles on low-frequency modes and instability in dusty plasma [6], You *et al.* made the measurement and analysis of the electron-neutral collision frequency in the calibration cut-off probe [7], Pecseli *et al.* analyzed the influence of neutral collision on the instability of current-driven electrostatic ion cyclotron [8], and Hahn *et al.* have modeled the collision process in the q-distributed plasma [9]. In addition, Khrapak derived the effective ion -neutral collision frequency in flowing plasma [10], Abbasi *et al.* calculated the collision time between plasma and the superthermal particles [11], Wang et al studied collision frequencies of particles in the plasma with power-law *q*-distribution in nonextensive statistics [12], Sun et al presented the average collision frequency of electron-neutral particle in the weakly ionized plasma with the power-law *q*-distribution [23] and Kuri *et al.* used mathematical model to simulate the impact of electron-ion collisions on the Weibel stability in the kappa-distributed unmagnetized plasma [13].

The research on the average collision frequency of plasma systems has always been one of the hot issues in plasma physics. The collision phenomenon is one of the most basic characteristics of gas kinetics in plasmas, Transport phenomenon in plasmas is usually related to collision effects [14-23], such as the diffusion, the heat conduction and migration and the particle drift in a magnetic field, etc. According to the degree of ionization in a plasma system, it can be divided into weakly ionized plasma, partially ionized plasma and fully ionized plasma. In a fully ionized plasma system, usually we think that electrons and ions are completely separated out and so the collisions of the plasma system are mainly taken place between the charged particles [14]. In a weakly ionized plasma system [15-23], there are neutral particles, ions and electrons, so there are electron-neutral particle collisions, ion-neutral particle collisions, and extremely few charged particle collisions such as electron-ion collisions and electron-electron collisions and ion-ion collisions. Since the collisions between electrons and neutral particles are the main kinetic mechanism for the weakly ionized plasma, it is of great significance to study the average collision frequency of electron-neutral particle collisions [7].

In nonextensive statistical mechanics, the particle velocity distribution function is the non-Maxwellian distribution or the power-law q-distribution, which describes the complex plasma being in a non-equilibrium state [24-27]. In many situations for astrophysical and space plasmas, the velocity distributions appear reasonably Maxwellian at low energies but have a "superthermal" power-law tail at high energies. In 1968, Vasyliunas introduced a non-Maxwellian empirical function to simulate the velocity distribution of high-energy electrons in the plasma magnetic sheath, which is the so-called kappa-distribution (or Vasyliunas distribution) [42]. The spacecraft measurements of the plasma velocity distribution in the solar wind, planetary magnetosphere and magnet field have showed that non-Maxwellian distributions are very common. In physics, astronomy, chemistry, biology and even social systems, non-Maxwellian distributions or power-law distributions are ubiquitous [28-30]. Therefore, the theoretical and experimental works of these distributions and their applications have attracted great attention in various of fields such as astronomy and astrophysics [31-35], plasmas and space physics [21,28,29,36-38], and reaction rate theory in chemistry [39-41], etc.

There are many space plasma phenomena, such as interstellar medium, thermo-ionosphere ionosphere, solar wind and planetary magnetosphere and magnet field, which reveal the universality of non-Maxwellian velocity distribution functions [42-45]. In 1995, Cairns *et al.* introduced a non-thermal velocity distribution function for electrons (or called the Carins distribution function) to characterize the existence of cavitons observed by Freja satellite [46] and Viking satellite [47]. In order to better study the linear and nonlinear phenomena in the complex plasma, Abid *et al.* gave a more general non-Maxwellian distribution function ( or Vasyliunas-Carins distribution function) in 2015 [48]. Recently, a generalized two-parameter ($r, q$) non-Maxwellian distribution function was studied and explored [49-53], which fits the CLUSTER data related to electrons in the magnetic sheath very well, and it is also effective for the solar wind proton data [54,55]. The two-parameter ($r, q$) distribution is very suitable for modeling the electron velocity distributions observed in the downstream region of the ground bow shock [53], and it can also be used in the general case of space plasma [50]. In 2017, Abid *et al.* further introduced a three-parameter velocity distribution function, including many non-Maxwellian distributions such as the Carins distribution function and the ($r, q$)-distribution function [53] etc. with spectral indices $\alpha$, $r$ and $q$ [56], which showed the rate of energetic and superthermality on the tail of velocity distribution curve of the plasma species, respectively. The three-parameter distribution function may help to understand the occurrence of a large number of linear and nonlinear space plasma phenomena in solar wind [57], magnetocaloric [55] and auroral regions [43,58] more effectively. The collision frequency between electron-neutral particles in the weakly ionized plasma plays an important role in these linear, nonlinear and transport phenomena. In order to study the kinetics of electrons in the non-Maxwellian velocity distribution functions of the nonequilibrium complex plasmas, in this work we study the average collision frequency of electrons-neutral-particles in the weakly ionized plasma with the non-Maxwellian distribution functions, and analyze the roles of these parameters in the average collision frequency of electron-neutral particles.

The paper is organized as follows. In section 2, we present several different types of non-Maxwellian/power-law distribution functions. In section 3, we study the average collision frequency between electrons and neutral particles in the weakly ionized plasmas with different non-Maxwellian distribution functions. In section 4, we numerically analyze the roles of these spectral indices in the average collision frequency. Finally in section 5, we give the conclusion.

## 2. Non-Maxwellian velocity distribution functions

Traditionally, the velocities of the $j$-th species in the plasma areconsidered to obey the Maxwellian velocity distribution when the plasma is in the thermal equilibrium state [59,60],

$$f_j^M(v_j) = \frac{n_j}{\pi^{3/2} v_{tj}^3} \exp(-v_j^2/v_{tj}^2), \tag{1}$$

where subscript $j$ represents the plasma species ($j = e$ is electrons and $j = i$ is ions), $n_j$, $v_j$, $T_j$, and $m_j$ are respectively the number density, the velocity, the temperature, and the mass of the particles, and $v_{tj} = (2k_B T_j / m_j)^{1/2}$ is the thermal velocity. This kind of distribution function (1) is only suitable for studying the plasma being in a thermal equilibrium state. Both experimental and theoretical studies have shown that when a complex plasma system deviates far from the thermal equilibrium state, the velocity distributions of particles can have a low-energy or superthermality tail shoulder [61-67]. In the recent years, non-Maxwellian velocity distributions are widely observed and studied in astrophysical and space plasmas and also in laboratory plasmas [68]. Next we will discuss several non-Maxwellian distribution functions.

In the astrophysical and space plasmas, the kappa distribution function (or Vasyliunas distribution function [42]) of the $j$-th species particle can be expressed as

$$f_j^V(v_j) = C_\kappa \frac{n_j}{\pi^{3/2} \kappa_1^3 v_{tj}^3} \left(1 + \frac{v_j^2}{\kappa \kappa_1^2 v_{tj}^2}\right)^{-(1+\kappa)}, \quad \kappa > \frac{3}{2}, \tag{2}$$

where the parameter $\kappa$ represents the superthermality in the plasma system, which shows the deviation of the complex space plasma from a thermal equilibrium; $\kappa_1 = \sqrt{(\kappa - 3/2)/\kappa}$, and $C_\kappa = \Gamma(1+\kappa)/\kappa^{3/2}\Gamma(\kappa - 1/2)$, where $\Gamma$ represents a gamma function. Only when we take the limit $\kappa \to \infty$, the Maxwellian distribution function is recovered.

The Carins distribution function [43] of velocity for the $j$-th species in the complex plasma can be written as follows,

$$f_j^C(v_j) = \frac{n_j}{\alpha_1 \pi^{3/2} v_{tj}^3} \left(1 + \alpha \frac{v_j^4}{v_{tj}^4}\right) \exp\left(-\frac{v_j^2}{v_{tj}^2}\right), \tag{3}$$

where the parameter $\alpha$ characterizes the number of high-energy species in the plasma system under consideration, and $\alpha_1 = 1 + 3\alpha$. Only if we take $\alpha = 0$, the Carins distribution becomes into the Maxwellian velocity distribution.

The Vasyliunas-Carins distribution function [48] of the $j$-th species in the complex plasma can be expressed as

$$f_j^{VC}(v_j) = \frac{n_j B_\kappa}{\pi^{3/2} v_{tj}^3} \left(1 + \alpha \frac{v_j^4}{v_{tj}^4}\right) \left(1 + \frac{v_j^2}{\kappa \kappa_1^2 v_{tj}^2}\right)^{-(1+\kappa)}, \tag{4}$$

where $B_\kappa$ is the normalization constant, given by

$$B_\kappa = \frac{1}{\kappa_1^{3/2} \kappa^{3/2}} \frac{\Gamma(1+\kappa)}{(1+3\alpha)\Gamma(\kappa - \frac{1}{2})}, \tag{5}$$

the two parameters need to satisfy $\kappa > 3/2$ and $\alpha < 1$. It is easily found that the Vasyliunas-Carins distribution function (4) can become (i) the Vasyliunas distribution function (2) only for $\alpha = 0$, (ii) the Carins distribution function (3) only for $\kappa \to \infty$, and (iii) the Maxwellian distribution function (1) only for $\alpha = 0$ and $\kappa \to \infty$.

The generalized two-parameter $(r, q)$ distribution function of the $j$-th species in the complex plasma can be written as

$$f_j^{r,q}(v_j) = D_{r,q} \left[1 + \frac{1}{q-1} \left(\frac{V_j^2}{X_{r,q} v_{tj}^2}\right)^{1+r}\right]^{-q}, \tag{6}$$

where $D_{r,q}$ is the normalization constant that depends on the parameters $r$ and $q$, given by

$$D_{r,q} = \frac{3 E_{r,q} n_j}{4\pi (X_{r,q})^{\frac{3}{2}} v_{tj}^3}, \tag{7}$$

where
$$X_{r,q} = \frac{3\Gamma\left(\frac{3}{2r+2}\right) \Gamma\left(q - \frac{3}{2r+2}\right)}{(q-1)^{\frac{1}{r+1}} \Gamma\left(q - \frac{5}{2r+2}\right) \Gamma\left(\frac{5}{2r+2}\right)}, \tag{8}$$

$$E_{r,q} = \frac{(q-1)^{\frac{-3}{2r+2}} \Gamma(q)}{\Gamma\left(q - \frac{3}{2r+2}\right) \Gamma\left(1 + \frac{3}{2r+2}\right)}. \tag{9}$$

The two-parameter $(r, q)$ distribution function (6) can show a broad shoulder with high energy tails. The deviation of the complex plasma from the thermal equilibrium depends significantly on the two parameters $r$ and $q$ [56]. It can be shown that when wetake $r = 0$ and $q = \kappa + 1$, the $(r, q)$-distribution function (6) becomes the kappa distribution function [42], and when we take $r = 0$ and $q \to \infty$, (6) becomes the Maxwellian distribution function. Based on the normalization and the definition of temperature in the above distribution function, the spectral index $r$ and $q$ must satisfy $q > 1$ and $q(1 + r) > 5/2$. Generally, if we increase the value of $r$ only and keep the value of $q$ unchanged, the contribution of high-energy particles decreases, but the shoulder of the distribution curve become wider. Similarly, if we increase the value of $q$ only and keep the value of $r$ fixed, the result is the same [69].

The three-parameter ($\alpha$, $r$, $q$) velocity distribution function (or the AZ distribution) of the $j$-th species in the complex plasma can be written as

$$f_j^{AZ}(v_j) = Y_{AZ}\left(1+\alpha\frac{v_j^4}{v_{tj}^4}\right)\left[1+\frac{1}{q-1}\left(\frac{v_j^2}{X_{r,q}v_{tj}^2}\right)^{r+1}\right]^{-q}, \tag{10}$$

where the normalization constant is modified [70] as

$$Y_{AZ} = \left(\frac{3n_j}{4\pi v_{tj}^3}\right)\frac{\rho_{r,q,\alpha}}{X_{r,q}^{3/2}}$$

with
$$\rho_{r,q,\alpha} = \frac{\Gamma(q)}{(1+9\eta_{r,q}\alpha)(q-1)^{\frac{3}{2r+2}}\Gamma\left(q-\frac{3}{2r+2}\right)\Gamma\left(1+\frac{3}{2r+2}\right)} \tag{11}$$

with
$$\eta_{r,q} = \frac{\Gamma\left(q-\frac{3}{2r+2}\right)\Gamma\left(\frac{3}{2r+2}\right)\Gamma\left(q-\frac{7}{2r+2}\right)\Gamma\left(\frac{7}{2r+2}\right)}{\Gamma^2\left(q-\frac{5}{2r+2}\right)\Gamma^2\left(\frac{5}{2r+2}\right)}. \tag{12}$$

The AZ-distribution function (10) has three parameters $\alpha$, $r$ and $q$, to satisfy the condition $q > 1$, $\alpha > 0$ and $q(1 + r) > 5/2$. They show the rate of charged particles on the shoulder, on a broad shoulder and the superthermality on the tail of velocity distribution curve of the plasma species, respectively. The three-parameter ($\alpha$, $r$, $q$) distribution function (10) is produced from the Carins distribution (3) and the two-parameter ($r$, $q$) distribution function (6). It is easily proved that when we take different values of the three-parameter ($\alpha$, $r$, $q$), the AZ-distribution function (10) can fall back to different non-Maxwellian velocity distribution functions. For example, when we take $\alpha = 0$, it becomes the ($r$, $q$) distribution function (6), when we take $r \to 0$ and $q = \kappa + 1$, it becomes the Vasyliunas-Cairns distribution function (4), when we take $r \to 0$ and $q \to \infty$, it becomes the Cairns distribution function (3), when we take $\alpha \to 0$, $r \to 0$ and $q = \kappa + 1$, it becomes the kappa distribution function (2), and only when we take $\alpha \to 0$, $r \to 0$ and $q \to \infty$, it recovers to the Maxwellian distribution function (1).

## 3. The average collision frequency of electron-neutral particle in the weakly ionized plasmas

The weakly ionized plasma is the plasma with an ionization degree of less than 1%, so the plasma temperature is about $(0.5\sim3.0)\times10^4$K [15-23]. In the weakly ionized plasma, the number of charged particles (electrons, ions) is relatively few. Therefore, when studying the collision of electrons, we usually only consider the collision between electrons and neutral particles [23]. Assuming that the kinetic energy of the electron is smaller or much smaller than the excitation energy of the neutral particle, the collision between the two can be regarded as elastic collision. If we regard electrons and neutral particles as hard spheres with no internal structure, when two particles collide elastically, there is no interaction force between the particles except the collision moment. Since the mass of an electron is much smaller than the mass of a neutral particle, the neutral particles can hardly obtain kinetic energy from the collision process with the electrons. Therefore, we can use the "steel ball" collision model [2] to calculate the average collision frequency of the electron-neutral particle in the weakly ionized plasma with the different types of non-Maxwellian distribution functions. The average collision frequency of the electron-neutral particle in this model is defined [2] as

$$\bar{v}_{en} = N_n \int v \sigma_{en}(v) f_e(\mathbf{v}) d\mathbf{v} , \qquad (13)$$

where $N_n$ is the number density of neutral particles, $\sigma_{en}$ is the collision scattering cross section, $f_e(\mathbf{v})$ is the velocity distribution function of electrons. We can easily find that the average collision frequency of the electron-neutral particle depends on the velocity distribution function. In the current plasma, we regard the electrons and the neutral particles as rigid spheres with radii $r_e$ and $r_n$ respectively. For the moving electrons and stationary neutral particles, collisions can occur when the projections of their centroids fall within a circular section with a radius of $d_{en}$, where $d_{en}$ is called the effective collision radius, which is equal to the sum of the radii of an electron and a neutral particle. Using the steel ball collision model, the collision scattering cross section can be expressed by

$$\sigma_{en} = \pi (r_n + r_e)^2 , \qquad (14)$$

where $r_e$ and $r_n$ are both constants for the plasma system under consideration, (13) becomes

$$\bar{v}_{en} = N_n \sigma_{en} \int v f_e(\mathbf{v}) d\mathbf{v} . \qquad (15)$$

By using equation (15) we can calculate the average collision frequency in the weakly ionized plasma following the velocity distribution $f_e(\mathbf{v})$. If the plasma follows the Maxwellian velocity distribution (1), the average collision frequency of the electron-neutral particle can be obtained as

$$\bar{v}_{en,M} = \frac{2\pi v_{te}}{\pi^{3/2}} N_n \sigma_{en} , \qquad (16)$$

where we have used the approximate conditions, $r_e \ll r_n$, $n_e \ll N_n$ and $\sigma_{en} = \pi(r_n + r_e)^2 \approx \pi r_n^2$. When the complex plasma under consideration deviates far from the equilibrium state, the velocity distribution of electrons is not a Maxwellian distribution, but often non-Maxwellian distribution.

If the velocity distribution of electrons in the weakly ionized plasma is the Vasyliunas distribution function in equation (2) (or the kappa distribution function), the average collision frequency of the electron-neutral particle can be derived as

$$\bar{v}_{en,V} = 4\pi^2 \frac{1}{\pi^{3/2}} \frac{\Gamma(1+\kappa)}{\kappa^{3/2}\Gamma(\kappa-\frac{1}{2})} \left(\frac{2\kappa}{2\kappa-3}\right)^{3/2} \frac{(3-2\kappa)^2}{8\kappa(\kappa-1)} v_{te} N_n r_n^2 . \tag{17}$$

We find that the above result significantly depends on the parameter $\kappa$. It is easy to prove that when the parameter $\kappa$ tends to infinity, the average collision frequency (17) returns to the result of that with a Maxwellian distribution. Similarly, when the velocity distribution of electrons obeys the Carins distribution function (3), the average collision frequency of the electron-neutral particle is calculated as

$$\bar{v}_{en,C} = \frac{1+6\alpha}{1+3\alpha} N_n r_n^2 \sqrt{\frac{8\pi k_B T_e}{m_e}}. \tag{18}$$

Obviously, when the nonthermal index $\alpha = 0$, the result of equation (18) is the same as the average collision frequency of the electron-neutral particle with a Maxwellian velocity distribution.

When the velocity distribution of electrons in the plasma is the Vasyliunas-Carins distribution function (4), the average collision frequency of the electron-neutral particle is given (see Appendix A) by

$$\bar{v}_{en,VC} = N_n r_n^2 \left(\kappa - \frac{3}{2}\right)^{\frac{5}{4}} \sqrt{\frac{2\pi k_B T_e}{m_e}} \frac{3\alpha(2\kappa-3)^2 + 2(\kappa-3)(\kappa-2)}{\kappa^{7/4}(1+3\alpha)(\kappa-3)(\kappa-2)(\kappa-1)} \frac{\Gamma(1+\kappa)}{\Gamma(\kappa-\frac{1}{2})}. \tag{19}$$

The condition that equation (19) holds is that the two spectral indices must satisfy $\kappa > 3$. It is clear that when we take $\kappa \to \infty$, equation (19) returns to the result of that with the Carins distribution, when we take $a = 0$ it returns to the result of that with the kappa distribution, and when we take $\kappa = \infty$ and $\alpha = 0$, the equation (19) will fall back to the result of a Maxwellian distribution in equation (16).

If the velocity distribution of electrons in the complex plasma can be simulated by the generalized two-parameter ($r$, $q$) distribution function in equation (6), the average collision frequency of the electron-neutral particle is derived (see Appendix B) as

$$\bar{v}_{en,(r,q)} = \sqrt{3}\pi N_n r_n^2 \sqrt{\frac{2k_B T_e}{m_e}} \left[\Gamma\left(\frac{3}{2r+2}\right)\Gamma\left(q-\frac{3}{2r+2}\right)\Gamma\left(q-\frac{5}{2r+2}\right)\Gamma\left(\frac{5}{2r+2}\right)\right]^{-\frac{1}{2}}$$
$$\times \Gamma\left(\frac{2}{1+r}\right)\Gamma\left(q-\frac{2}{1+r}\right). \tag{20}$$

where the parameters $q(1+r) > 5/2$ and $q > 1$ are required so as to ensure the convergence of the integral. It is worth noting that under the limits of $r = 0$ and $q \to \infty$, equation (20) will reduce to the case of a Maxwellian velocity distribution.

If the velocity distribution function of electrons can be depicted with the AZ distribution function (10), the average collision frequency of the electron-neutral particle is calculated (see Appendix C) as

$$\bar{v}_{en,AZ} = \frac{A_{r,q,\alpha}\pi r_n^2 N_n \sqrt{6 k_B T_e/m_e}}{\Gamma^2\left(q-\frac{5}{2r+2}\right)\Gamma^2\left(\frac{5}{2r+2}\right) + 9\alpha\Gamma\left(q-\frac{3}{2r+2}\right)\Gamma\left(\frac{3}{2r+2}\right)\Gamma\left(q-\frac{7}{2r+2}\right)\Gamma\left(\frac{7}{2r+2}\right)},$$

(21)

where

$$A_{r,q,\alpha} = 9\alpha \Gamma\left(\frac{4}{1+r}\right)\Gamma\left(q-\frac{4}{1+r}\right)\frac{\left[\Gamma\left(\frac{3}{2r+2}\right)\Gamma\left(q-\frac{3}{2r+2}\right)\right]^{3/2}}{\left[\Gamma\left(q-\frac{5}{2r+2}\right)\Gamma\left(\frac{5}{2r+2}\right)\right]^{1/2}}$$
$$+ \Gamma\left(\frac{2}{1+r}\right)\Gamma\left(q-\frac{2}{1+r}\right)\frac{\left[\Gamma\left(q-\frac{5}{2r+2}\right)\Gamma\left(\frac{5}{2r+2}\right)\right]^{3/2}}{\left[\Gamma\left(\frac{3}{2r+2}\right)\Gamma\left(q-\frac{3}{2r+2}\right)\right]^{1/2}}, \quad (22)$$

and the convergence of the integral requires that $q > 1$, $\alpha > 0$ and $q(r+1) > 5/2$. Equation (21) shows the evolution of $v_{en,AZ}$ with the parameters $\alpha$, $r$ and $q$. From equation (21) we can obtain the conclusions that (i) when we take $\alpha = 0$, (21) becomes into $v_{en,(r,q)}$ in equation (20); (ii) when we take $r \to 0$ and $q = \kappa + 1$, (21) becomes into $v_{en,VC}$ in equation (19); (iii) when we take $r \to 0$ and $q \to \infty$, (21) becomes into $v_{en,C}$ in equation (18); (iv) when we take $r \to 0$, $\alpha \to 0$ and $q = \kappa + 1$, (21) becomes into $v_{en,V}$ in equation (17); (v) when we take $\alpha \to 0$, $r \to 0$ and $q \to \infty$, (21) becomes into $v_{en,M}$ in equation (16).

## 4. Numerical analysis and discussion

In section 3, it has been shown that the spectral indices in the non-Maxwell distribution functions play an important role in the average collision frequency of the electron-neutral particle in the weakly ionized complex plasma. Now by numerical analyses we demonstrate how these spectral indices, or the non-Maxwellian velocity distributions, influence the properties of the average collision frequency. For this purpose, by using equation (17), equation (18) and equation (16), we can write the following equations,

$$\frac{\bar{v}_{en,V}}{\bar{v}_{en,M}} = \frac{\sqrt{2}}{2}\frac{\Gamma(1+\kappa)}{\Gamma(\kappa-\frac{1}{2})}\frac{(2\kappa-3)^{\frac{1}{2}}}{\kappa(\kappa-1)}, \text{ and } \frac{\bar{v}_{en,C}}{\bar{v}_{en,M}} = \frac{1+6\alpha}{1+3\alpha}. \quad (23)$$

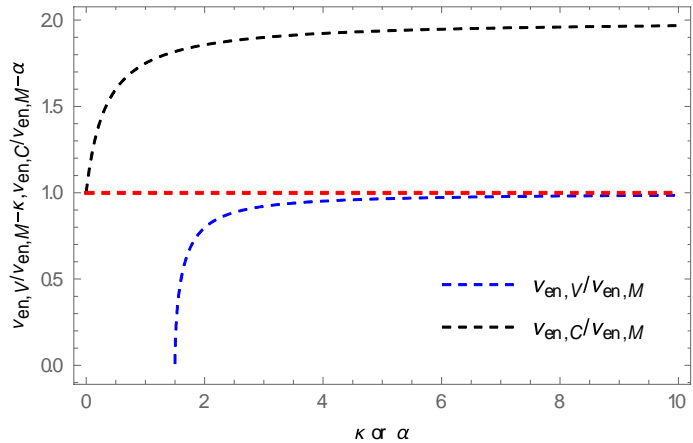

Figure1. Dependence of the average collision frequency in (17) on the parameter $\kappa$ and in (18) on the parameter $\alpha$, respectively.

The results obtained from equation (23) are displayed in figure 1, where the red dashed line represents $\bar{v}_{en,M}/\bar{v}_{en,M}$, which is the standard 1, and the blue dashed line represents the ratio of the average collision frequency of the electrons-neutral-particle collision in the weakly ionized plasma with the Vasyliunas velocity distribution (kappa distribution) to that with the Maxwellian distribution, and the black dashed line represents the ratio of the average collision frequency with the Carins velocity distribution to that with the Maxwellian distribution. Figure1 clearly shows that $\bar{v}_{en,V}$ increases as the parameter $\kappa$ increases for low values of $\kappa$ (viz., $\kappa < 7$). As $\kappa$ continues to increase to a certain value (viz., $\kappa = 10$), it will gradually approach the standard curve "$\bar{v}_{en,M}/\bar{v}_{en,M}$", and it reaches the maximum value in the limit of $\kappa \to \infty$. On the other hand, $\bar{v}_{en,C}$ is always greater than $\bar{v}_{en,M}$ and $\bar{v}_{en,V}$, and it takes the minimum value when $\alpha = 0$. After that, as $\alpha$ continues to increase, $\bar{v}_{en,C}$ shows a trend of rapid increase within the small value of $\alpha$, and then $\bar{v}_{en,C}$ does not vary basically with the $\alpha$ increase (viz., $\alpha > 7$).

From equations (19) and (16), we find that

$$\frac{\bar{v}_{en,VC}}{\bar{v}_{en,M}} = \left(\kappa - \frac{3}{2}\right)^{5/4} \frac{\left[3\alpha(2\kappa-3)^2 + 2(\kappa-3)(\kappa-2)\right]}{2(1+3\alpha)\kappa^{7/4}(\kappa-3)(\kappa-2)(\kappa-1)} \frac{\Gamma(1+\kappa)}{\Gamma(\kappa-\frac{1}{2})}. \quad (25)$$

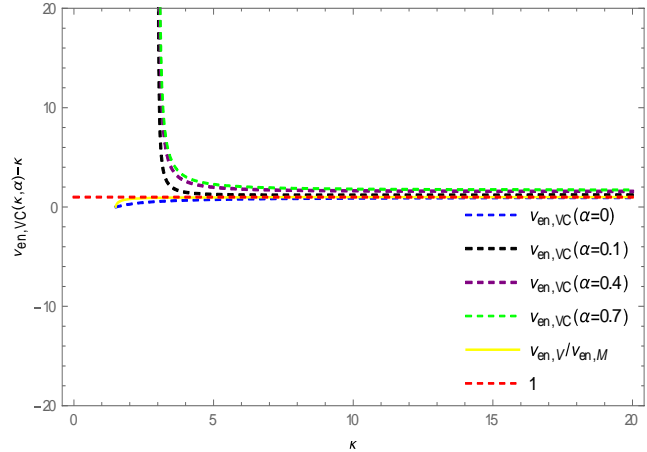

(a)

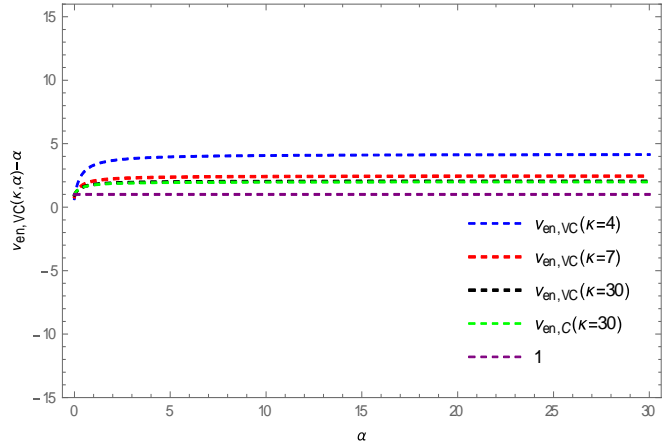

(b)

Figure 2. Dependence of the average collision frequency in (19) on parameters $\kappa$ and $\alpha$.

Based on equation (25), in figures 2 (a) and (b), we show the effects of the parameters $\kappa$ and $\alpha$ on the average collision frequency (19) of the electron-neutral particle in the plasma with Vasyliunas-Carins distribution. Numerically analyses of $\bar{v}_{en,VC}$ are made for different values of $\alpha$ (viz., $\alpha = 0$, $\alpha = 0.1$, $\alpha = 0.4$ and $\alpha = 0.7$ ) and different values of $\kappa$ (viz., $\kappa = 4$, $\kappa = 6$ and $\kappa = 9$ ). It is observed from figure 2(a) that $\bar{v}_{en,VC}$ decrease sharply with the increase of $\alpha$ within a small range of $\kappa$ (viz., $\kappa < 8$) for a fixed $\alpha$, and when $\kappa$ exceeds a certain range, such as $\kappa > 8$, $\bar{v}_{en,VC}$ will approach $\bar{v}_{en,M}$. It is worth noting that when $\alpha = 0$, the curve of $\bar{v}_{en,VC}$ is similar to that of $\bar{v}_{en,V}$. We also show the dependence of $\bar{v}_{en,VC}$ curve on the spectral index $\alpha$ for a fixed $\kappa$ in figure 2(b), which indicates that $\bar{v}_{en,VC}$ intersects the $\bar{v}_{en,M}$ curve when $\alpha = 0$, and when $\alpha$ is small and in a certain range (viz., $\alpha < 5$), it increases monotonously with the increase of $\alpha$. When $\alpha$ exceeds a certain value, $\bar{v}_{en,VC}$ is close to a constant. Another property presented in Figure 2(b) is that as $\kappa$ increases, $\bar{v}_{en,VC}$ will gradually approach $\bar{v}_{en,C}$, and in the limit of $\kappa \to \infty$ the two coincide completely.

From equations (20) and (16), we can find that

$$\frac{\bar{v}_{en,(r,q)}}{\bar{v}_{en,M}} = \frac{\sqrt{3\pi}}{2}\left[\Gamma\left(\frac{3}{2r+2}\right)\Gamma\left(q-\frac{3}{2r+2}\right)\Gamma\left(q-\frac{5}{2r+2}\right)\Gamma\left(\frac{5}{2r+2}\right)\right]^{-\frac{1}{2}}\Gamma\left(\frac{2}{1+r}\right)\Gamma\left(q-\frac{2}{1+r}\right). \quad (26)$$

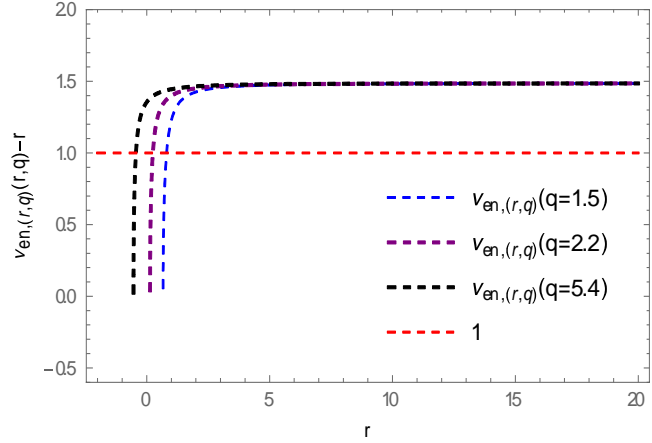

(a)

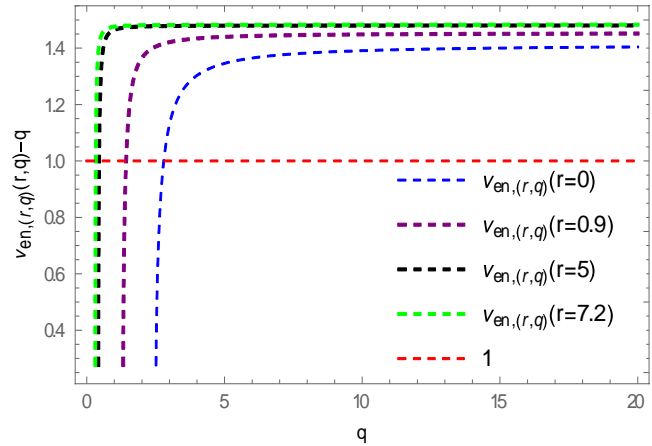

(b)

Figure 3. Dependence of the average collision frequency in equation (20) on the parameters $r$ and $q$.

Based on equation (26), figure 3 numerically shows the effects of spectral indices $r$ and $q$ on the average collision frequency in equation (20). In figure 3(a) we took $q$ as three fixed values of 1.5, 2.2 and 5.4, respectively, and then changed the parameter $r$. It can be seen that as the value of $r$ increases, $\bar{v}_{en,(r,q)}$ increases for the low values of $r$ (viz., $r < 5$), and for large values of $r$ with different $q$ values, all $\bar{v}_{en,(r,q)}$ will approach a same constant. In figure 3(b), $r$ was taken a fixed as 0, 0.9, 5 and 7.2, respectively, and the values of $q$ changed. We notice that $\bar{v}_{en,(r,q)}$ increases rapidly as the value of $q$ increases when $q$ is small, but it becomes constant when value of $q$ is higher and it becomes independent of $q$. And with the increase of $r$, in the limit of $q \to \infty$ all $\bar{v}_{en,(r,q)}$ will approach the same constant. For this reason, we can give a mathematical expression by calculating the limit of equation (26) at $r \to \infty$, as follows:

$$\lim_{r \to \infty} \frac{\bar{v}_{en,(r,q)}}{\bar{v}_{en,M}} = \frac{3}{8} \sqrt{\frac{5\pi \Gamma^2(q)}{\Gamma(q^2)}} . \qquad (27)$$

Based on equations (21) and (16), the dependence of $\bar{v}_{en,AZ}$ on the spectral indices $\alpha$, $r$ and $q$ is presented by

$$\frac{\bar{v}_{en,AZ}}{\bar{v}_{en,M}} = \frac{1}{2} \frac{\sqrt{3\pi} A_{r,q,\alpha}}{\Gamma^2\left(q - \frac{5}{2r+2}\right)\Gamma^2\left(\frac{5}{2r+2}\right) + 9\alpha \Gamma\left(q - \frac{3}{2r+2}\right)\Gamma\left(\frac{3}{2r+2}\right)\Gamma\left(q - \frac{7}{2r+2}\right)\Gamma\left(\frac{7}{2r+2}\right)} . \qquad (28)$$

In order to illustrate our the results more clearly, in figure 4(a)-(c), based on equation (28) we show the dependence of the average collision frequency $\bar{v}_{en,AZ}$ on the parameters $\alpha$, $r$ and $q$. From the figure 4 (a) we show that $\bar{v}_{en,AZ}$ decreases with the increase of the parameter $q$ when $\alpha$ and $r$ are fixed and $\alpha \neq 0$. When $q$ increases to a certain value (related to the value of $r$ and $\alpha$), $\bar{v}_{en,AZ}$ will become a constant and independent of $q$. This is because $\bar{v}_{en,AZ}$ transforms into $\bar{v}_{en,(r,q)}$ at $\alpha = 0$, the curve of $\bar{v}_{en,AZ}$ is also infinitely close to that of $\bar{v}_{en,(r,q)}$, as shown in figure 4(a), the black dotted line and the green dotted line are completely merged.

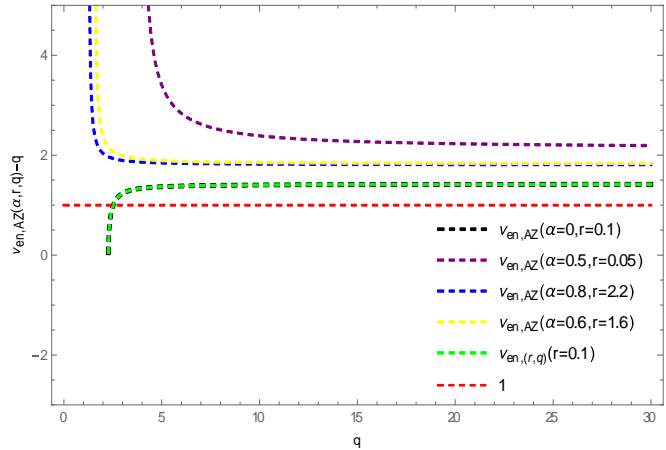

(a)

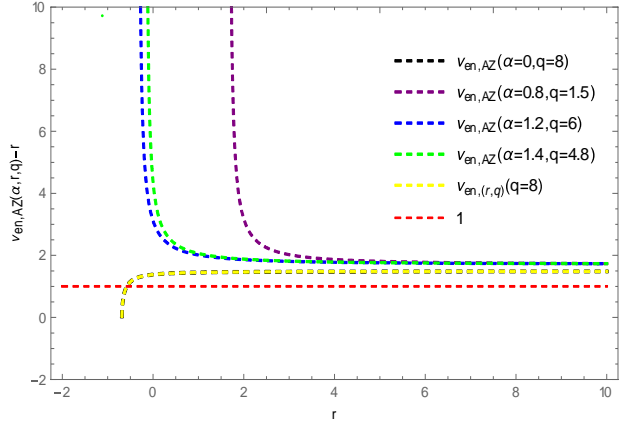

(b)

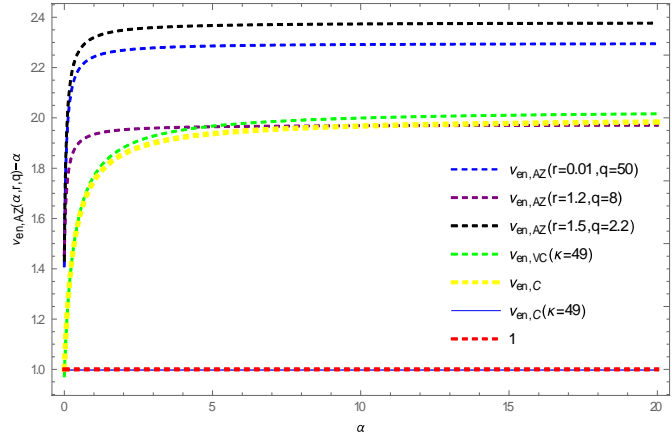

(c)

Figure 4. Dependence of the average collision frequency in equation (21) on the parameters $\alpha$, $r$ and $q$.

Figure 4(b) depicts that for fixed values of $\alpha$ ($\alpha \neq 0$) and $q$ (viz., $\alpha = 0.8$ and $q = 1.5$, $\alpha = 1.2$ and $q = 6$), the frequency $\bar{v}_{en,AZ}$ decreases with increase of the parameter $r$, and when $r$ reaches a certain value, $\bar{v}_{en,AZ}$ no longer changes with change of $r$. From the black and yellow curves in figure 4(b) it can be seen that $\bar{v}_{en,AZ}$ will fall back to $\bar{v}_{en,(r,q)}$ for $\alpha = 0$. In figure 4(c) we show that $\bar{v}_{en,AZ}$ depends on $\alpha$ significantly when $r$ and $q$ are fixed. In this case, $\bar{v}_{en,AZ}$ increases rapidly with the increase of $\alpha$ in the range $0 < \alpha < 5$, and at $\alpha = 0$, $\bar{v}_{en,AZ}$ takes the minimum value. When $\alpha$ is greater than a certain value (viz., $\alpha > 8$), $\bar{v}_{en,AZ}$ will hardly change with $\alpha$. In all cases, we find that $\bar{v}_{en,M}$ is generally smaller than $\bar{v}_{en,AZ}$.

## 5. Conclusion and Discussion

In summary, we have investigated the average collision frequency of electron-neutral-particle in the weakly ionized complex plasmas with a variety of non-Maxwellian velocity distributions. We derived the expressions of the average collision frequency of electron-neutral-particle when the plasmas follow the kappa distribution, the Carins distribution, the Vasyliunas Carins distribution, the generalized two parameter ($r, q$) distribution, and the three-parameters AZ velocity distribution, respectively. They are given by the equations (17), (18), (19), (20) and (21), respectively. We find that these average collision frequencies are all significantly dependent on the parameters in the complex plasmas with the non-Maxwell distributions. And only when these spectral indices (parameters) are the special cases, the collision frequencies in equations (17), (18), (19), (20) and (21) can all return to the cases of the plasmas with the Maxwellian distribution perfectly.

In order to show the results more clearly, we have numerically analyzed the effects of the non-Maxwellian distributions on the average collision frequency of electron-neutral-particle in the weakly ionized complex plasmas. The results are illustrated in the figures (1)-(4) respectively.

We have shown that the frequency $\bar{v}_{en,V}$ in the plasma is generally smaller than $\bar{v}_{en,M}$, and $\bar{v}_{en,V}$ increases monotonously with the increase of the parameter $\kappa$ at first and finally it approaches $\bar{v}_{en,M}$ with increase of $\kappa$. The frequency $\bar{v}_{en,C}$ depends strongly on $\alpha$ when $\alpha$ is relatively small, and such dependence will be weaken gradually with the increase of $\alpha$.

We have also shown that when $\alpha$ is fixed, the frequency $\bar{v}_{en,VC}$ has a sharp decrease as the increase of the parameter $\kappa$ in a small range of $\kappa$, and then $\bar{v}_{en,VC}$ hardly changes with $\kappa$. For a fixed value of $\kappa$, $\bar{v}_{en,VC}$ increases sharply at the beginning and then approaches a constant gradually.

We have further shown the effects of $r$ and $q$ on the frequency $\bar{v}_{en,(r,q)}$. It is found that $\bar{v}_{en,(r,q)}$ increases as $r$ increase when value of $r$ is low, but when $r$ exceeds a certain range, all $\bar{v}_{en,(r,q)}$ will approach the same constant. We can get the similar conclusion that the difference is that $\bar{v}_{en,(r,q)}$ with different fixed $r$ values will approach different constants when the value of $q$ becomes high.

Finally, we have respectively studied the roles of the parameters $\alpha$, $r$ and $q$ in the average collision frequency of electron-neutral-particle in the complex plasma with the three-parameter AZ velocity distribution. If the parameters $\alpha$ and $r$ are fixed, we show that when value of $q$ is low, the frequency $\bar{v}_{en,AZ}$ decrease monotonously with the increase of $q$ and finally it approaches a constant. If the parameters $\alpha$ and $q$ are fixed, we show that the effect of $r$ on $\bar{v}_{en,AZ}$ is similar to the effect of $q$ on $\bar{v}_{en,AZ}$. But if $r$ and $q$ are fixed, $\bar{v}_{en,AZ}$ increase monotonously with the increase of $\alpha$ when $\alpha$ is small. When $\alpha$ exceeds a certain value, $\bar{v}_{en,AZ}$ will also become a constant. We can obtain the following points:

(i) for $\alpha = 0$, $\bar{v}_{en,AZ}$ becomes $\bar{v}_{en,(r,q)}$.
(ii) for $r \rightarrow 0$ and $q = \kappa + 1$, $\bar{v}_{en,AZ}$ becomes $\bar{v}_{en,VC}$.
(iii) for $\alpha \rightarrow 0$ and $r \rightarrow 0$ and $q = \kappa + 1$, $\bar{v}_{en,AZ}$ becomes $\bar{v}_{en,V}$.
(iv) for $r \rightarrow 0$ and $q \rightarrow \infty$, $\bar{v}_{en,AZ}$ and $\bar{v}_{en,C}$ become the same.
(v) for $\alpha \rightarrow 0$, $r \rightarrow 0$ and $q \rightarrow \infty$, $\bar{v}_{en,AZ}$ becomes $\bar{v}_{en,M}$.

In conclusion, the average collision frequencies of electron-neutral-particle in the weakly ionized complex plasmas with the non-Maxwellian velocity distributions are generally and significantly different from those with a Maxwellian velocity distribution, and so the related transport properties in the complex plasmas are also different from those in the plasmas with a Maxwellian velocity distribution.

**Appendix A**

The calculation of $\bar{v}_{en,VC}$. Substituting equation (4) into equation (15), we obtain that

$$\bar{v}_{en,VC} = N_n \sigma_{en} \int v f_e(\mathbf{v}) d\mathbf{v}$$

$$= 4\pi \frac{n_e B_\kappa}{\pi^{3/2} v_{te}^3} N_n \sigma_{en} \int_0^\infty v^3 \left(1 + \alpha \frac{v^4}{v_{te}^4}\right)\left(1 + \frac{2}{2\kappa - 3}\frac{v^2}{v_{te}^2}\right)^{-(1+\kappa)} dv$$

$$= 4\pi \frac{n_e B_\kappa}{\pi^{3/2} v_{te}^3} N_n \sigma_{en} \frac{v_{te}^4 (3-2\kappa)^2 \left[3\alpha(3-2\kappa)^2 + 2(\kappa-3)(\kappa-2)\right]}{16(\kappa-3)(\kappa-2)(\kappa-1)\kappa}$$

$$= N_n r_n^2 \sqrt{\frac{2\pi k_B T_e}{m_e}} \left(\kappa - \frac{3}{2}\right)^{\frac{5}{4}} \frac{3\alpha(2\kappa-3)^2 + 2(\kappa-3)(\kappa-2)}{(1+3\alpha)\kappa^{7/4}(\kappa-3)(\kappa-2)(\kappa-1)} \frac{\Gamma(1+\kappa)}{\Gamma(\kappa-\frac{1}{2})}. \quad (A1)$$

where $\kappa > 3$ is required to ensure the convergence of the integral.

**Appendix B**

The calculation of $\bar{v}_{en,(r,q)}$. We can express the integral of the equation (15) that obeys the distribution function (6) as,

$$\bar{v}_{en,(r,q)} = N_n \sigma_{en} \int v f_e(\mathbf{v}) d\mathbf{v} = 4\pi D_{r,q} N_n \sigma_{en} \int_0^\infty v^3 \left(1 + \frac{1}{q-1}\left(\frac{v^2}{X_{r,q} v_{te}^2}\right)^{1+r}\right)^{-q} dv$$

$$\approx \pi^2 D_{r,q} N_n r_n^2 (q-1)^{\frac{2}{1+r}} \left(\frac{1}{v_{te}^2 X_{r,q}}\right)^{-2} \frac{\Gamma\left(\frac{3+r}{1+r}\right)\Gamma\left(q - \frac{2}{1+r}\right)}{\Gamma(q)}$$

$$= \frac{3\pi}{4} N_n r_n^2 \frac{n_e E_{r,q}}{\Gamma(q)} v_{te} (X_{r,q})^{1/2} (q-1)^{\frac{2}{1+r}} \frac{2}{1+r} \Gamma\left(\frac{2}{1+r}\right) \Gamma\left(q - \frac{2}{1+r}\right)$$

$$= \frac{3\sqrt{3}}{4} \pi N_n r_n^2 n_e v_{te} \left[\frac{\Gamma\left(\frac{3}{2r+2}\right)\Gamma\left(q - \frac{3}{2r+2}\right)}{\Gamma\left(q - \frac{5}{2r+2}\right)\Gamma\left(\frac{5}{2r+2}\right)}\right]^{\frac{1}{2}} \frac{4}{3} \frac{\Gamma\left(\frac{2}{1+r}\right)\Gamma\left(q - \frac{2}{1+r}\right)}{\Gamma\left(q - \frac{3}{2r+2}\right)\Gamma\left(\frac{3}{2r+2}\right)}$$

$$\approx \pi N_n r_n^2 \sqrt{\frac{6 k_B T_e}{m_e}} \left[\Gamma\left(\frac{3}{2r+2}\right)\Gamma\left(q - \frac{3}{2r+2}\right)\Gamma\left(q - \frac{5}{2r+2}\right)\Gamma\left(\frac{5}{2r+2}\right)\right]^{-\frac{1}{2}} \Gamma\left(\frac{2}{1+r}\right)\Gamma\left(q - \frac{2}{1+r}\right). \quad (B2)$$

where $q(1+r) > 5/2$ and $q > 1$ are required to ensure the convergence of the integral.

**Appendix C**

The calculation of $\bar{v}_{en,AZ}$ is calculated as follows.

$$\bar{v}_{en,AZ} = N_n \sigma_{en} \int v f_e(\mathbf{v}) d\mathbf{v} = 4\pi Y_{AZ} N_n \sigma_{en} \int_0^\infty v^3 \left(1 + \alpha \frac{v^4}{v_{te}^4}\right) \left(1 + \frac{1}{q-1}\left(\frac{v^2}{X_{r,q} v_{te}^2}\right)^{r+1}\right)^{-q} dv$$

$$= \pi N_n \sigma_{en} Y_{AZ} \frac{(q-1)^{\frac{4}{1+r}} (X_{r,q})^4 v_{te}^4}{2\Gamma(q)} \left[\alpha \Gamma\left(\frac{5+r}{1+r}\right)\Gamma\left(q-\frac{4}{1+r}\right) + \frac{2(X_{r,q})^{-2}}{(q-1)^{\frac{2}{1+r}}} \Gamma\left(\frac{3+r}{1+r}\right)\Gamma\left(q-\frac{2}{1+r}\right)\right]$$

$$= N_n \sigma_{en} n_e \frac{3 v_{te} (q-1)^{\frac{4}{1+r}} (X_{r,q})^{\frac{5}{2}} \rho_{r,q,\alpha}}{8\Gamma(q)} \left[\alpha \Gamma\left(\frac{5+r}{1+r}\right)\Gamma\left(q-\frac{4}{1+r}\right) + \frac{2(X_{r,q})^{-2}}{(q-1)^{\frac{2}{1+r}}} \Gamma\left(\frac{3+r}{1+r}\right)\Gamma\left(q-\frac{2}{1+r}\right)\right]$$

$$= \frac{N_n \sigma_{en} n_e v_{te} (q-1)^{\frac{5}{2r+2}} (X_{r,q})^{\frac{5}{2}}}{(1+9\eta_{r,q}\alpha)\Gamma\left(q-\frac{3}{2r+2}\right)\Gamma\left(\frac{3}{2r+2}\right)} \left[\alpha \Gamma\left(\frac{5+r}{1+r}\right)\Gamma\left(q-\frac{4}{1+r}\right) + \frac{2(X_{r,q})^{-2}}{(q-1)^{\frac{2}{1+r}}} \Gamma\left(\frac{3+r}{1+r}\right)\Gamma\left(q-\frac{2}{1+r}\right)\right]$$

$$= N_n \sigma_{en} n_e v_{te} (q-1)^{\frac{5}{2r+2}} \frac{\Gamma^2\left(q-\frac{5}{2r+2}\right)\Gamma^2\left(\frac{5}{2r+2}\right)}{\Gamma\left(q-\frac{3}{2r+2}\right)\Gamma\left(\frac{3}{2r+2}\right)}$$

$$\frac{\alpha(X_{r,q})^{\frac{5}{2}}\Gamma\left(\frac{4}{1+r}\right)\Gamma\left(q-\frac{4}{1+r}\right) + (q-1)^{-\frac{2}{1+r}}(X_{r,q})^{\frac{1}{2}}\Gamma\left(\frac{2}{1+r}\right)\Gamma\left(q-\frac{2}{1+r}\right)}{\Gamma^2\left(q-\frac{5}{2r+2}\right)\Gamma^2\left(\frac{5}{2r+2}\right) + 9\alpha\Gamma\left(q-\frac{3}{2r+2}\right)\Gamma\left(\frac{3}{2r+2}\right)\Gamma\left(q-\frac{7}{2r+2}\right)\Gamma\left(\frac{7}{2r+2}\right)}$$

$$\approx \frac{A_{r,q,\alpha} \pi r_n^2 N_n \sqrt{6 k_B T_e / m_e}}{\Gamma^2\left(q-\frac{5}{2r+2}\right)\Gamma^2\left(\frac{5}{2r+2}\right) + 9\alpha\Gamma\left(q-\frac{3}{2r+2}\right)\Gamma\left(\frac{3}{2r+2}\right)\Gamma\left(q-\frac{7}{2r+2}\right)\Gamma\left(\frac{7}{2r+2}\right)}.$$

(C3)

where $q > 1$, $\alpha > 0$ and $q(1+r) > 5/2$ are satisfied to ensure the validity of the integral.

## Acknowledgments

This work is supported by the National Natural science foundation of China under Grant No. 11775156.